\documentclass[10pt,twoside]{article}

\usepackage{7OSME-style}

\usepackage{float}
\usepackage{bm}
\usepackage{hyperref}
\usepackage{breakurl}

\newcommand{\hrefurl}[1]{\href{#1}{#1}}

\newcommand{\bA}{\bm{A}}
\newcommand{\bB}{\bm{B}}
\newcommand{\ba}{\bm{a}}
\newcommand{\bb}{\bm{b}}
\newcommand{\bq}{\bm{q}}
\newcommand{\bp}{\bm{p}}
\newcommand{\bx}{\bm{x}}

\newcommand{\chg}[1]{#1}

\title{The Cosmic Spiderweb and\\ General Origami Tessellation Design}
\authorlist[Neyrinck]{Mark C.\ Neyrinck} 
\affiliations{Mark C.\ Neyrinck \at University of the Basque Country, Department of Theoretical Physics, 48080 Bilbao, Spain; Ikerbasque Fellow. \email{Mark.Neyrinck@gmail.com} }

\makeatletter
  \let\runtitle\@title
  \let\runauthor\shortauthor
\makeatother

\begin{document}

\maketitle

\begin{abstract}
The cosmic web (the arrangement of matter in the universe), spider's webs, and origami tessellations are linked by their geometry (specifically, of sectional-Voronoi tessellations). This motivates origami and textile artistic representations of the cosmic web. It also relates to the scientific insights origami can bring to the cosmic web; we show results of some cosmological computer simulations, with some origami-tessellation properties. We also adapt software developed for cosmic-web research to provide an interactive tool for general origami-tessellation design.
\end{abstract}

\section{Introduction}
\label{sec:intro}

The `cosmic web' is an influential concept in modern physical cosmology, the term introduced in the paper `How filaments of galaxies are woven into the cosmic web' \cite{BondEtal1996}. This paper showed that if gravity acts as expected neighboring clusters of galaxies should generally be joined by filaments of small galaxies, like pearls on a necklace; also, neighboring small galaxies should be linked by filaments of gas and dark matter. The concept essentially existed in the 70's and 80's in the work of Zeldovich and collaborators, but it took observations such as the CfA redshift survey `stick figure' \cite{deLapparentEtal1986} and the term `cosmic web' to firmly establish the idea in cosmology.

Tom\'{a}s Saraceno\footnote{See e.g.\ {\it How to Entangle the Universe in a Spider Web}, {\it 14 Billions} and {\it Galaxies forming along filaments, like droplets along the strands of a spider's web} at \hrefurl{http://www.tomassaraceno.com}} \cite{Ball2017} has noticed visual similarities between the cosmic web and spider's webs, constructing human-scale enlargements of black-widow spider's webs. \cite{DiemerFacio2017} review these and other textile artistic representations, and describe new ways of constructing them.

In a recent paper \cite{NeyrinckEtal2018}, we provided a rigorous explanation of the correspondence between the cosmic web and spiderwebs, also linking both to origami tessellations. In the next section, we discuss several ingredients that led us to this result, defining terms along the way.

\section{From a spider's web to the cosmic web}
\label{sec:links}
The first step is from a spider's web (an actual web spun by a spider) to a `spiderweb,' a structural-engineering term for a spatial network of nodes joined by edges that can be strung up to be entirely in tension. Threads in spider's webs can sag. But often, these threads could be pulled taut, to satisfy the spiderweb condition. Previous OSME papers \cite{LangBateman2011,Lang2015} showed the rigorous link between spiderwebs and origami tessellations, discussed in section \ref{sec:origamidesign}.

\subsection{Spiderwebs}
Geometrically, spiderwebs are defined in terms of dual graphs. Consider a planar (with non-crossing edges, in 2D) spatial graph with positioned nodes, and straight edges linking them; call this the {\it primal} graph. This graph tessellates the plane into polygonal, non-overlapping cells. Define a set of {\it generators}, one per cell. Construct the {\it dual} graph by drawing edges connecting neighboring generators (with neighboring cells). If the edge connecting each pair of generators is perpendicular to the edge separating the generators' cells, the primal and dual graphs are {\it reciprocal}. The primal graph is a {\it spiderweb} if the edges of the dual tessellation do not cross each other.

The concept of a reciprocal dual was introduced by James Clerk Maxwell \cite{Maxwell1864}, much better known for uniting electricity and magnetism. He used reciprocal duals to analyze and design pin-jointed structural-engineering trusses. This grew into the subject of graphic statics, used to build many structures in the late 1800's, e.g.\ the Eiffel Tower. Graphic statics waned somewhat in popularity in the 20th century, but remained a topic of study, e.g.\ in Walter Whiteley's structural topology group, which coined the term `spiderweb' around 1980 (Ethan Bolker, private communication). With modern computer techniques and visualization, there are several groups actively studying and using fully 3D graphic statics \cite{McRobie2016,KonstantatouMcRobie2016,McRobie2017,BlockEtal2016,AkbarzadehEtal2016}.

\begin{SCfigure}
    \centering
    \includegraphics[width=0.5\linewidth]{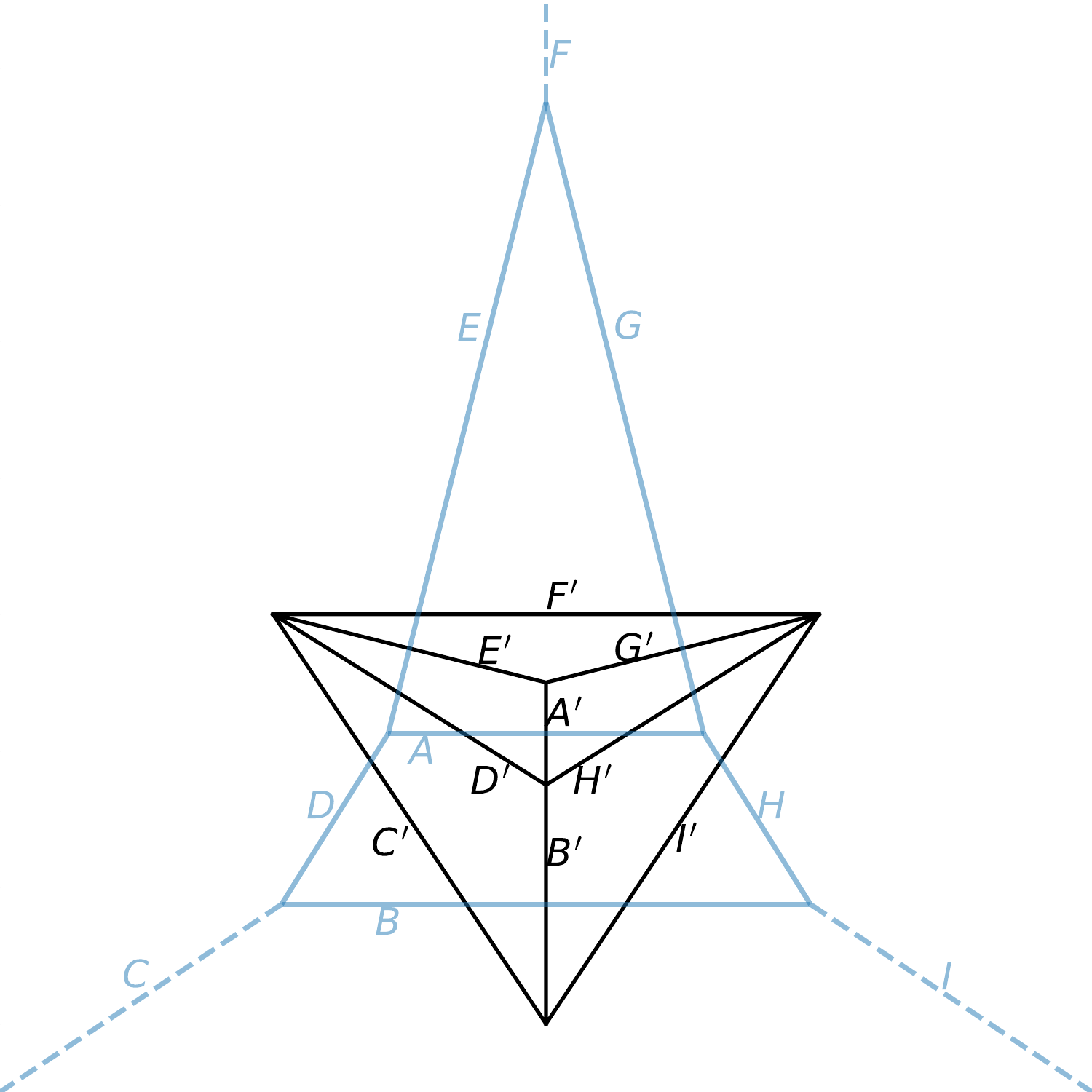}
  \caption{A spiderweb form diagram (grey) resembling the Eiffel Tower, and the corresponding force diagram (black). Letters label perpendicular pairs of form (unprimed) and force (primed) edges. Some perpendicular segment pairs would only actually intersect if extended. Dashed edges are external supports.}
  \label{fig:eiffeltower}
\end{SCfigure}

In structural engineering, the primal graph is called the {\it form diagram}, i.e.\ the map of structural members in a truss, with nodes and edges. The {\it force diagram} is the dual, reciprocal graph, so-named because the length of each edge is proportional to the force in that structural member. Maxwell showed that if and only if the network is in equilibrium, a closed force diagram can be constructed such that the form and force networks are reciprocals of each other. Figure \ref{fig:eiffeltower} shows an example of a spiderweb. The reciprocal-dual force diagram appears in black; the form diagram is a spiderweb because all force polygons are closed and fit together without crossing any edges. 

In Figure \ref{fig:eiffeltower}, the form diagram is a {\it Voronoi} tessellation, a partition of space into polygonal cells, one for each of a set of generating points. The cell for each generator is the patch of space closest to that generator. The force diagram is the dual, {\it Delaunay} tessellation, an edge drawn between each neighboring pair of generators. Here, the generators are vertices of the force diagram, in black. All Voronoi/Delaunay dual tessellations are reciprocal, since the Voronoi edges are perpendicular bisectors to the Delaunay edges. For clarity, the generators of both tessellations are coincident, but note that the force diagram (with different units than the form diagram) can be arbitrarily scaled and retain its reciprocal nature.

Voronoi/Delaunay duals give a wide class of spiderwebs. The entire class of spiderwebs is larger, but only a bit. Voronoi edges are perpendicular bisectors of their corresponding Delaunay edges; the `bisector' part can be relaxed. Each Voronoi edge may be slid along its Delaunay edge, closer to one of the generators. They may not be slid entirely independently, though, since the Voronoi edges must still join at vertices. There turns out to be one extra degree of freedom per generator, causing its cell to expand or contract.

The result is a {\it sectional-Voronoi} diagram, a section through a higher-dimensional Voronoi tessellation. A generator's extra degree of freedom in a sectional-Voronoi diagram can be thought of as its distance from the space being tessellated. A sectional-Voronoi diagram can also be thought of as a Voronoi tessellation in which each generator may have a different additive `power' in the distance function used to determine which points are closest to the generator (thus an alternative term, `power diagram'). \cite{AshBolker1986} showed that 2D spiderwebs and sectional-Voronoi tessellations are equivalent. See also the more informal \cite{WhiteleyEtal2013}.

\chg{In symbols, the sectional-Voronoi cell} $V_{\bq}$ around the generator at position $\bq$ is
\begin{equation}
V_{\bq}=\left\{\bx\in E~{\rm s.t.}~ |\bx-\bq|^2 + z_q^2 \le |\bx-\bp|^2 + z_p^2,~\forall \bp\in L\right\},
\label{eqn:secvoronoi}
\end{equation}
where $E$ is the space being tessellated, $L$ is the set of generators \chg{within $E$}, and $z_q$ and $z_p$ are constants, possibly different at each point $\bp$ and $\bq$, and interpretable as distances of points of $L$ away from $E$ in a higher dimension. If all $z_p$ and $z_q$ are the same, the tessellation reduces to an ordinary Voronoi diagram. We call the dual, reciprocal tessellation a weighted-Delaunay tessellation; it is also known as a `regular tessellation.'

\subsection{From spiderwebs to the cosmic web}
Remarkably, the cosmic web is approximately a sectional-Voronoi tessellation, as well; since the cosmic web has a spiderweb geometry, we call it the `cosmic spiderweb.' The equivalence is exact in a good approximation, called the {\it adhesion model}, for how gravity arranges matter into the the cosmic web; below we describe this. 

On scales much larger than galaxies, after subtracting off the expansion of the universe, matter particles have hardly moved from their primordial location. It is a reasonable approximation to give each matter particle a push based on the pattern of primordial density around it, and then simply let the particle coast ballistically on its original trajectory \cite{Zeldovich1970}. This formalism breaks down most noticeably when trajectories cross, and particles blithely fly past each other; full gravity would make the particles switch direction, ultimately pulling them into a collapsed structure like a filament or galaxy.

The adhesion model \cite{GurbatovSaichev1984,KofmanEtal1990,GurbatovEtal2012} eliminates this over-crossing problem with a mechanism that sticks trajectories together when they cross. A viscosity is introduced formally into the equation of motion (resulting in a differential equation called Burgers' equation), and then the viscosity is reduced to 0. A few methods exist to solve for the resulting structure; most elegant, arguably, is a convex-hull construction \cite{VergassolaEtal1994}, which gives a sectional-Voronoi diagram \cite{HiddingEtal2012,HiddingEtal2016,NeyrinckEtal2018,HiddingEtal2018,Hidding2018}. 

In this construction, each cell $V_{\bq}$ is defined by Eq.\ \ref{eqn:secvoronoi}, with the weight at $\bq$ given by $z_q^2=-2\Phi(\bq)$. Here $\Phi(\bq)$ is the `displacement potential' governing how patches move from their initial positions. It is a `potential' as used in physics; in 1D, the displacement from initial position $q$ to final position $x$ is minus the derivative of the potential, $x-q=-d\Phi(q)/dq$; in 2D and 3D, this derivative becomes a gradient. Interestingly, this gradient can be taken with a sectional-Voronoi tessellation.

As the universe ages and expands, $\Phi(\bq)$ \chg{advances. To illustrate, we use the Zeldovich approximation (ZA), in which the entire potential scales uniformly with a multiplicative factor as time passes, starting at 0 at the Big Bang. The ZA is already quite accurate \cite{ColesEtal1993}, and the prescription for changing $\Phi$ with time can be improved upon rather simply \cite{Neyrinck2013,Neyrinck2016truthstretch}}.

In {\it comoving} coordinates (scaling out the mean expansion of the universe), patches on potential hills expand, and patches in potential wells contract. \chg{Generators remain at their initial positions within the space being tessellated (say, a 2D horizontal $x$-$y$ plane), but slide down with time to $z_q=-\sqrt{C-\Phi(\bq)}$, where $C$ is an arbitrary constant. The physically meaningful cosmic web is a section, at the $x$-$y$ plane, through the full 3D Voronoi tessellation. Cells in expanding potential hills stay near the plane, but cells in collapsing potential wells sink farthest away, their Voronoi cells eventually ceasing to intersect the plane at all; then, they have completely collapsed and disappeared from the cosmic web. All of this behavior is controlled by} the scalar $\Phi(\bq)$.

\begin{SCfigure}
    \centering
    \includegraphics[width=0.3\linewidth]{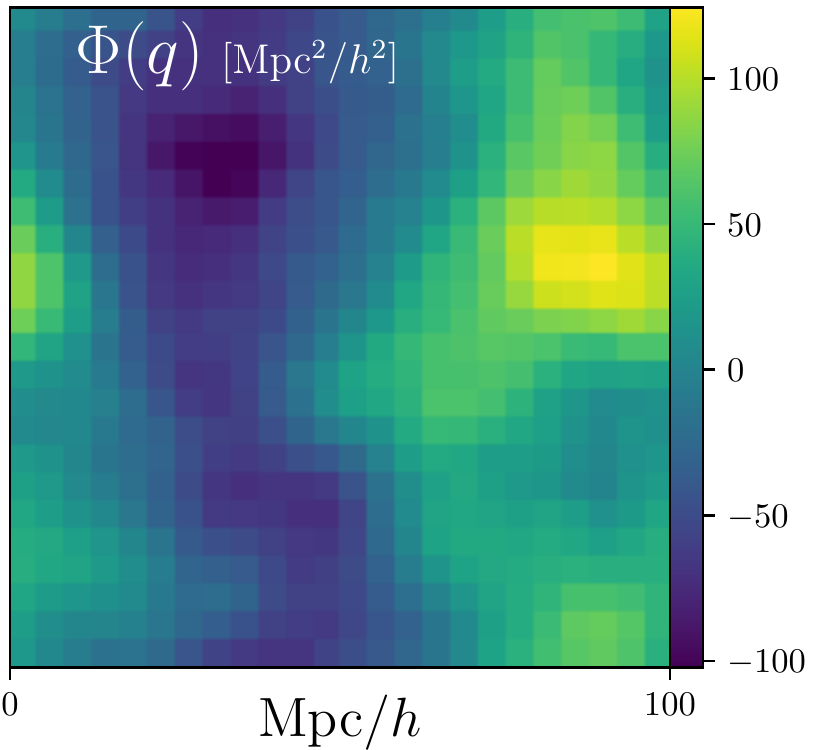}
  \caption{The displacement potential used to generate the following adhesion-model cosmic webs.}
     \label{fig:displacement_potential}
\end{SCfigure}

\begin{figure}
	\centering
    \includegraphics[width=\linewidth]{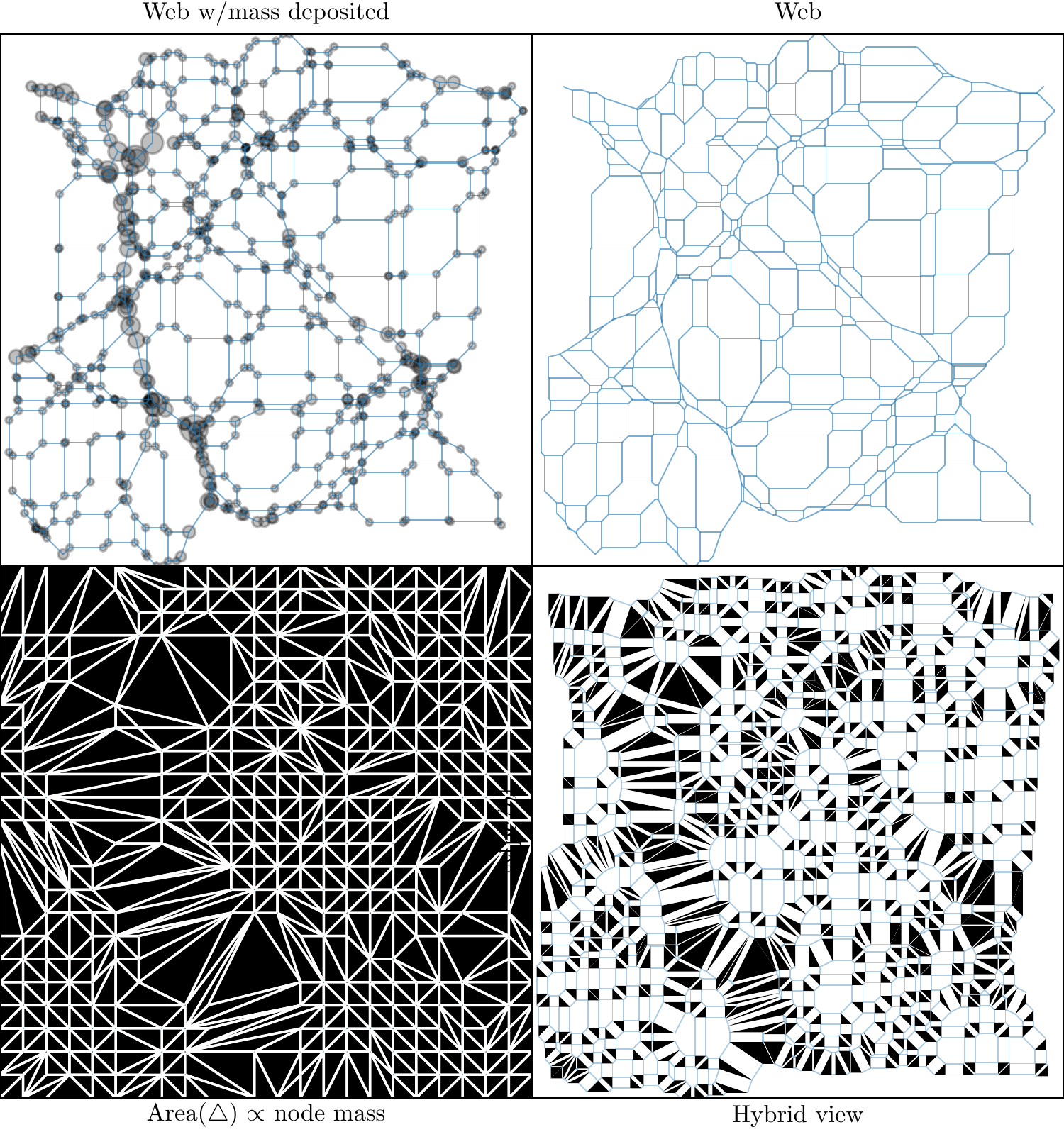}
  \caption{{\it Upper right}: A cosmic web generated from the displacement potential in Figure \ref{fig:displacement_potential}. Each white polygon bordered in grey is a sectional-Voronoi cell, inhabiting Eulerian (final position) space; the web collectively is a spiderweb. {\it Lower left}: In white, the corresponding reciprocal dual tessellation, in Lagrangian (initial comoving position) space; each node of the Eulerian web is a black triangle here. In architecture, the length of each white edge is proportional to the tension in the corresponding spiderweb thread.  {\it Upper left}: the web at upper right, adding a translucent black circle at each node, with area proportional to its mass (the area of its black triangle at lower left). {\it Lower right}: A Minkowski sum of the dual tessellations, every cell halved in linear size.}
    \label{fig:cosmicduals}
\end{figure}

A $\Phi(\bq)$ field appears in Figure \ref{fig:displacement_potential}, and the resultant cosmic spiderweb in Figure \ref{fig:cosmicduals}. Each triangle in initial, {\it Lagrangian} coordinates, with mass given by its area (lower left) is a node of the spiderweb in actual, {\it Eulerian} space (upper right). The nodes are shown with mass deposited at upper left, and a half-half Lagrangian/Eulerian mixture is at lower right, called a Minkowski sum, a construction resembling a crease pattern for an origami tessellation.

A {\it Minkowski sum} of two sets of vectors $\bA$ and $\bB$ is $\bA+\bB \equiv \{\ba+\bb~|~\ba\in\bA, \bb\in\bB\}$. We follow \cite{McRobie2016}, adapting the concept for reciprocal tessellations, in which there is a subset $\bB_i$ of the dual tessellation $\bB$ attachable to each vector $\ba_i\in\bA$. We also add an \chg{arbitrary} scaling $\alpha$ to interpolate between the original and dual tessellations. The vertices satisfy
\begin{equation}
\alpha\bA+(1-\alpha)\bB\equiv \left\{\alpha\ba_i+(1-\alpha)\bb_j~|~\ba_i\in\bA, \bb_j\in\bB_i\right\}.
\label{eqn:minksum}
\end{equation}
The vectors in the sum are what we plot.

\begin{figure}
  \centering
    	\includegraphics[width=0.6\columnwidth]{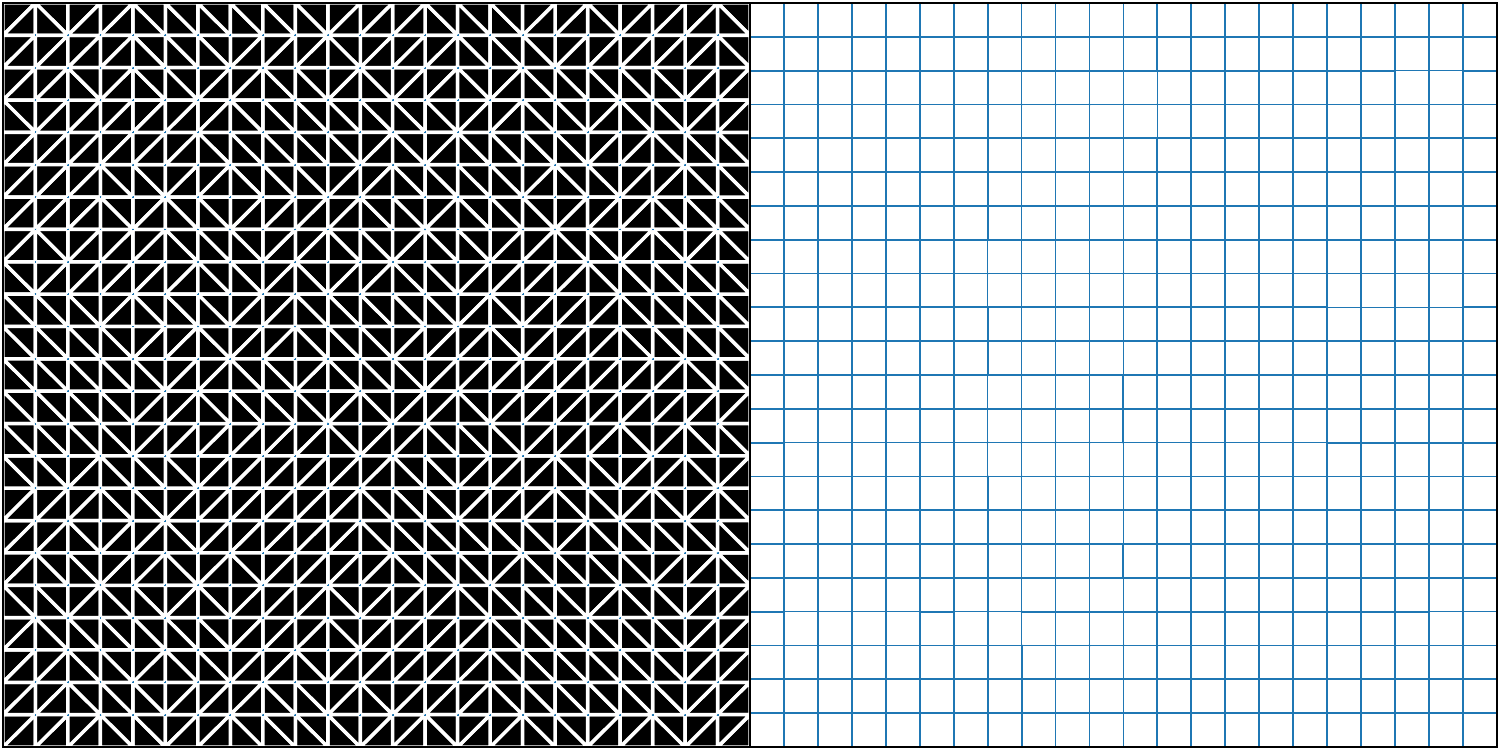}
    	\includegraphics[width=0.6\columnwidth]{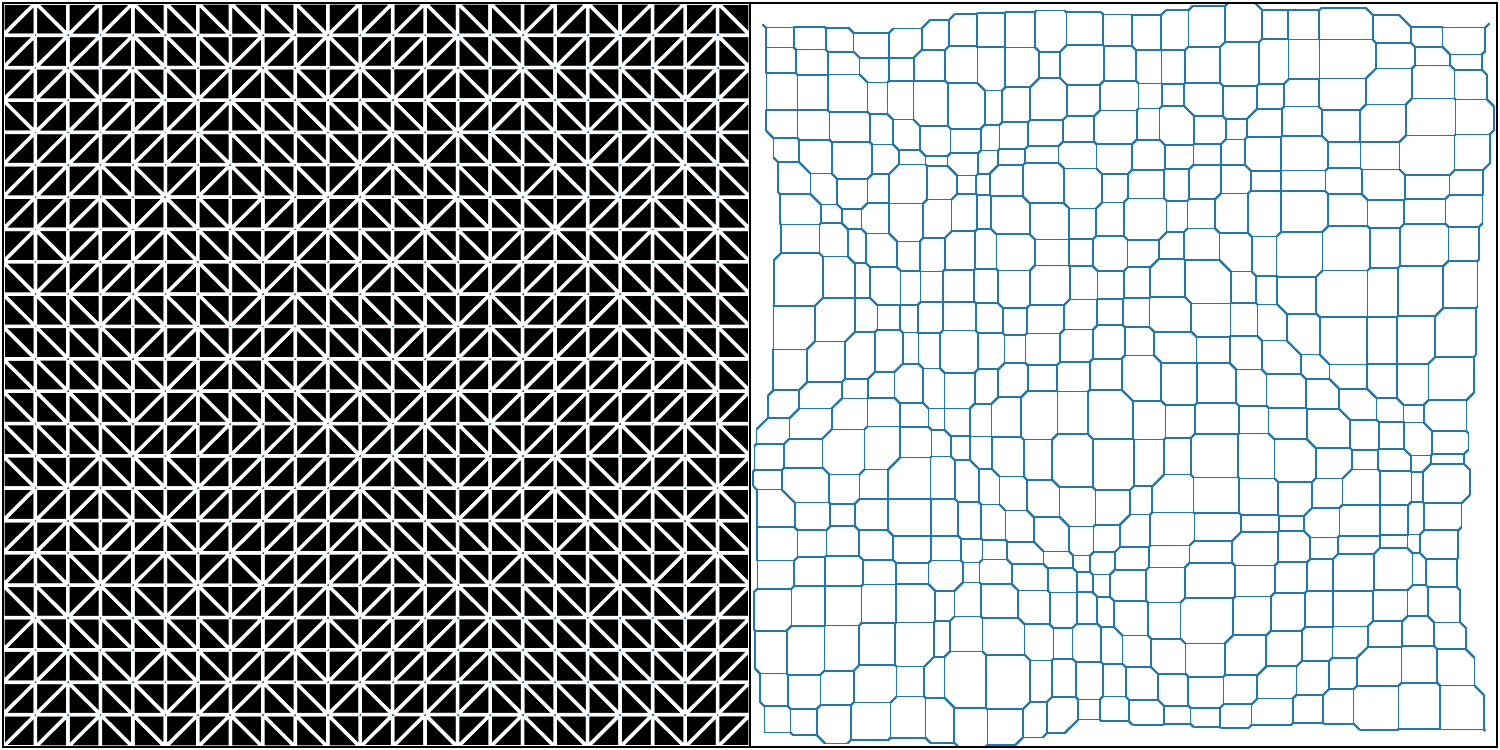}
    	\includegraphics[width=0.6\columnwidth]{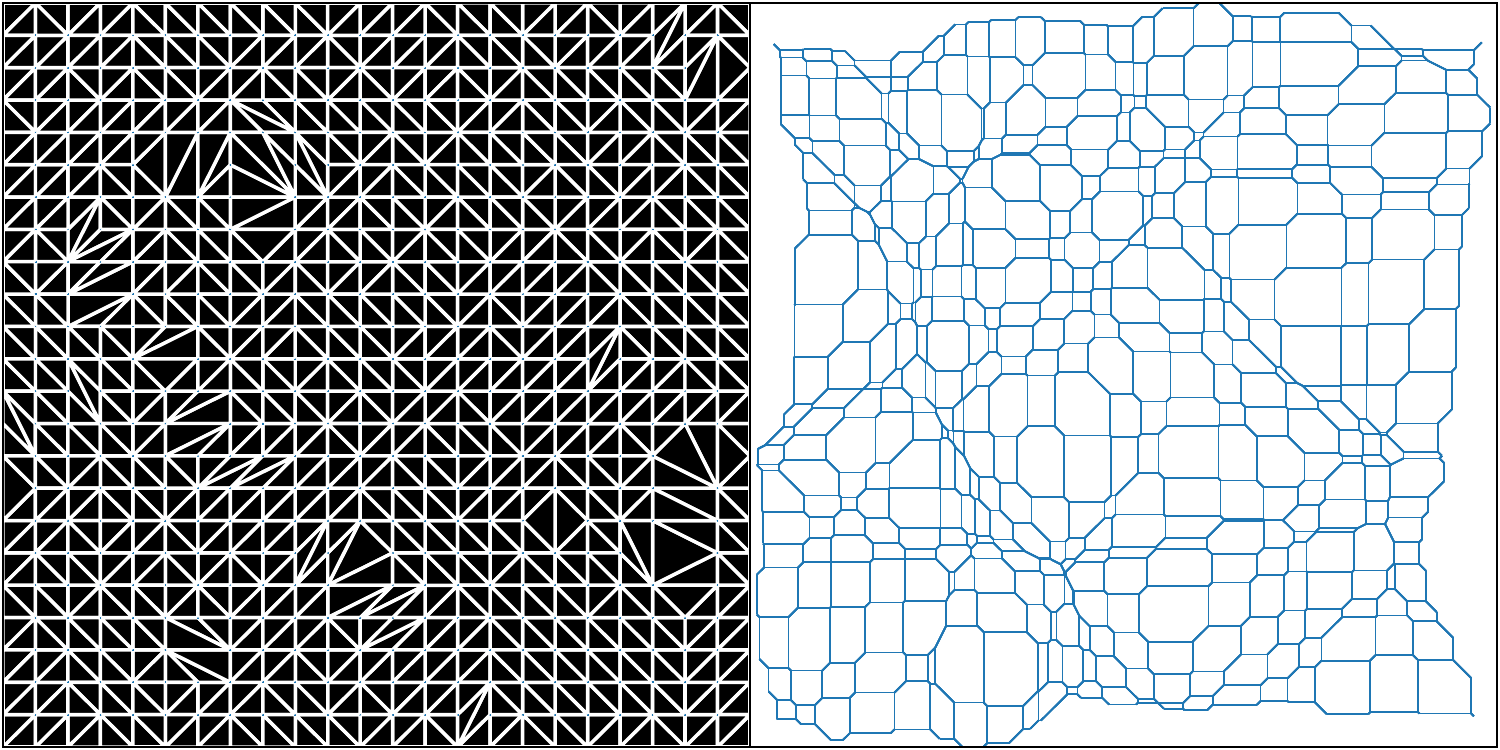}
    	\includegraphics[width=0.6\columnwidth]{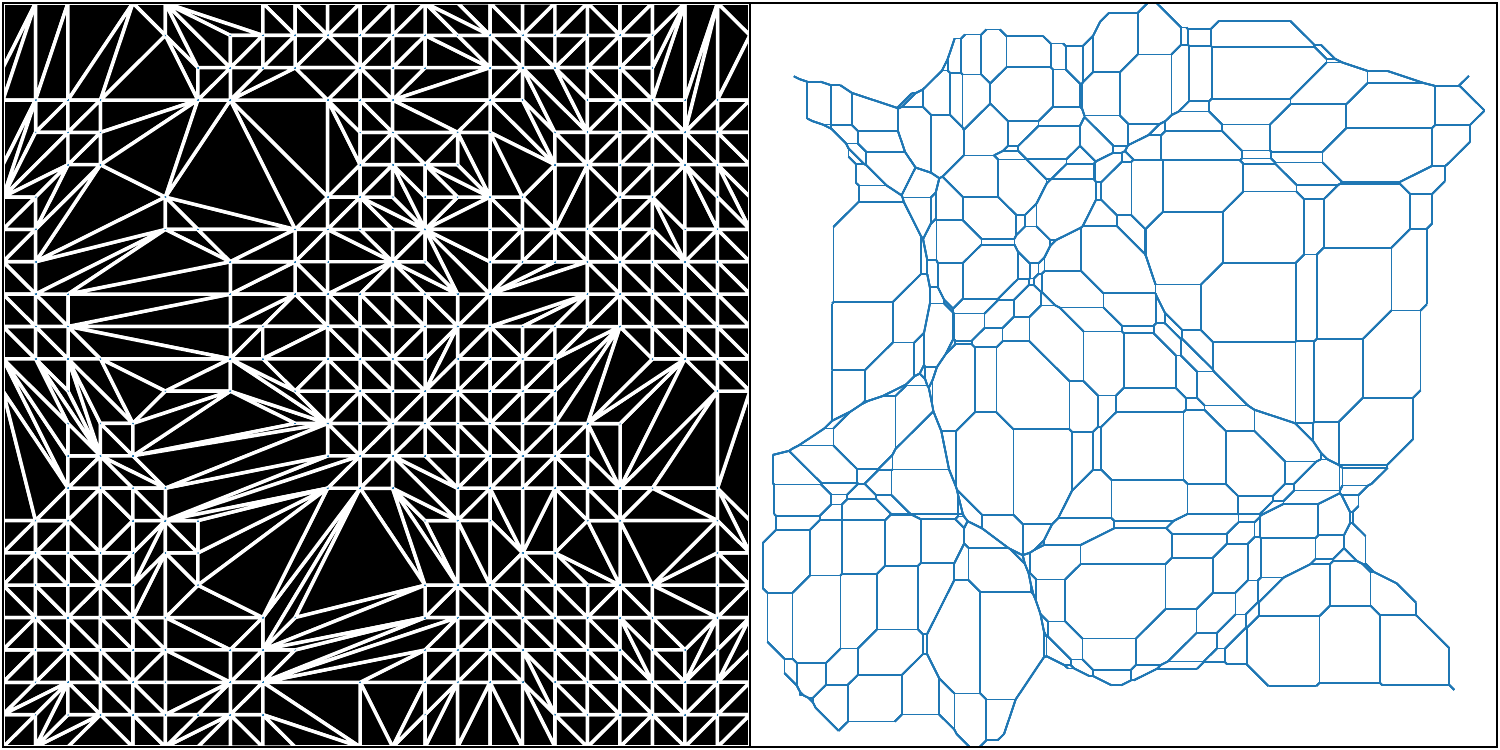}
  \caption{A time sequence of the adhesion-model cosmic web in Figure \ref{fig:cosmicduals}, scaling the displacement potential in Figure \ref{fig:displacement_potential} by $0$, $0.25$, $0.5$, and $1$, from top to bottom. {\it Left}: Dual triangulation; each black patch collapses into a node of the web at right.}
  \label{fig:timesequence}
\end{figure}

Figure \ref{fig:timesequence} shows the progression with time of this cosmic web, from uniformity with zero displacement, to the snapshot in Figure \ref{fig:cosmicduals}.

\subsection{From the adhesion model to reality}
We assert that the adhesion model is `good,' but what does this mean for the real universe? The `cosmic spiderweb' concept has some caveats:

\begin{itemize}
\item The adhesion model does not treat aspects of the real universe: rotational motions, and structure inside collapsed regions (such as groups, clusters, and filaments of galaxies). So, we should not expect structures inside the Milky Way, or even the Milky Way plus its satellite galaxies, to be spiderwebs. Fortunately for the approximation, cosmic expansion is thought to severely dampen rotational motion outside of collapsed regions, but some may persist; in fact, a possible use for the `cosmic spiderweb' concept is to test for unexpected rotation.

\item It is ambiguous how to translate actual observations of the universe to a form amenable to comparison with a structural-engineering spiderweb. But there are many different observational or theoretical classifications of the cosmic web that could be used \cite{LibeskindEtal2018}; progress can be made by choosing one, and taking into account any issues that arise.

\item For an exact spiderweb, cosmic-web nodes must be included not only in collapsed regions, but in uncollapsed voids (not included in usual conceptions of the cosmic web). These give kinks to filaments of the universe. In many cases, the departure from a spiderweb will be negligible if only nodes in galaxies and filaments are included, but this is something to test.

\item The real cosmic web has curves, built as it is from catastrophe theory \cite{Arnold1993,McRobie2017seduction}. This differs from the angular set of straight lines depicted here, because we show the cosmic web approximated at a coarse resolution; its full structure emerges only at high resolution.
\end{itemize}

While these caveats may present challenges, the `spiderwebiness' of the cosmic web is a new observable for testing theories of cosmology and structure formation. For instance, a departure from the perpendicularity property of a spiderweb could indicate that distances are estimated according to the wrong cosmological model. See \cite{NeyrinckEtal2018} for further discussion.

\subsection{Three dimensions}
We have concentrated on 2D here, because it is simpler, is straightforwardly visualized, and it is relevant to paper origami. But the universe is spatially 3D. Also, the field of fully 3D graphic statics has experienced a resurgence of interest, and is currently an active area of research.

Most of the spiderweb concepts we have discussed generalize straightforwardly to 3D. \cite{Rankine1876} introduced the concept of a 3D reciprocal dual. As in 2D, the form diagram is the map of a truss's members and nodes in space, and the force diagram is a collection of fitted-together force polyhedra, one polyhedron per node. Also, the edges of 3D sectional-Voronoi tessellations give spiderwebs, as in 2D. An ambiguity, though, is that a 3D tessellation has panels, as well as edges. They may have structural-engineering importance as well, but the spiderweb of edges is what have essentially the same properties as in 2D. Completing the `cosmic spiderweb' idea in 3D, the adhesion-model formalism produces a sectional-Voronoi tessellation, as well.

There is no fully rigorous presentation of the 3D spiderweb/sectional-Voronoi \chg{relationship in the literature as far as we know}, but the unpublished \cite{CrapoWhiteley1994} contains some essential concepts. Most importantly for the `cosmic spiderweb' idea, a sectional-Voronoi tessellation is a spiderweb. \chg{But unlike in 2D, in 3D the converse is not generally true}; for instance, \cite{McRobie2016} gives an example of a spiderweb 3D truss built from two glued-together trusses that are separately, but not together, sectional-Voronoi.

\section{Origami and the cosmic web}
The above link between spiderwebs, origami, and the cosmic web provides additional justification to previous work on cosmological origami \cite{Neyrinck6OSME2015}. There are reasons to explore what insights origami mathematics might bring to cosmology: like origami, the formation of structures in the cosmos is a folding process. It was an early application of catastrophe theory \cite{ArnoldEtal1982,HiddingEtal2014,FeldbruggeEtal2018}: a manifold occupies a higher-dimensional space, but is only straightforwardly observable in a lower dimension into which it is projected. In the spatially 3D universe, this manifold is a 3D `dark-matter sheet' \cite{ShandarinEtal2012,AbelEtal2012} that resides in 6D position-velocity phase space; it is most observable when projected back to 3D position space. Also, origami tessellations visually resemble cosmological density fields.

Although the term can encompass much more, in this paper, I mean by `origami tessellation' a simple flat twist tessellation. This is a flat-folding origami design comprised of three types of regions: {\it void} regions that may translate from the unfolded to folded state, but are not allowed to rotate; {\it filament} regions comprised of parallel pleat {\it caustics} (creases); and {\it node} regions that are simple twist folds.

Indeed, the cosmic web must form an origami tessellation in a strict, `origami' approximation, in which the density field can change only by folding and piling up layers of the dark-matter sheet, but not by stretching it; see \cite{Neyrinck6OSME2015}. This no-stretch approximation has little physical justification; in reality, the dark-matter sheet stretches substantially, although it remains curiously close to unstretched on average deep within galaxies \cite{VogelsbergerWhite2011}.

\chg{Arising from the well-motivated adhesion model, the `cosmic spiderweb' concept has more physical justification} than this origami approximation, which however is still useful as a toy model for the formation of the cosmic web. Particularly, the concept of an origami tessellation from a 3D `sheet' is intriguing. In 3D, creases are 2D surfaces. The twist fold is a concept that carries into 3D \cite{Neyrinck2016tetcol}; I call a 3D twist fold a `tetrahedral collapse' in a cosmological context. A 2D twist fold is a convex polygon that twists by some angle, generating a pleat from each \chg{edge} of the polygon. A 3D twist fold is a convex polyhedron that inverts through the center, and twists by a 3D angle. \chg{Each face of the polyhedron is a crease. As it twists, from each face of the polyhedron is extruded a filament (a tube whose cross section is that face); from each edge comes a wall (a parallel pair of polygons); and from each vertex comes a void (a polyhedron filling the space between neighboring twist folds). The directions of the filaments and the 3D twist angle suffice to determine all of these properties. See \cite{Neyrinck2016tetcol} for 3D figures and animations of this geometry.}

This model should be particularly useful in understanding how nearby galaxies tend to rotate similarly to each other, e.g.\ \cite{SlosarEtal2009}. This is because generally, galaxies rotate, and form with filaments that join them to each other. In tetrahedral collapse, nodes rotate if and only if filaments rotate as well; this implies that rotating filaments of darkness may be common in the universe, correlating the rotations of galaxies they connect.

\subsection{Gravitational simulations}
\begin{figure}
\centering
\includegraphics[width=0.496\linewidth]{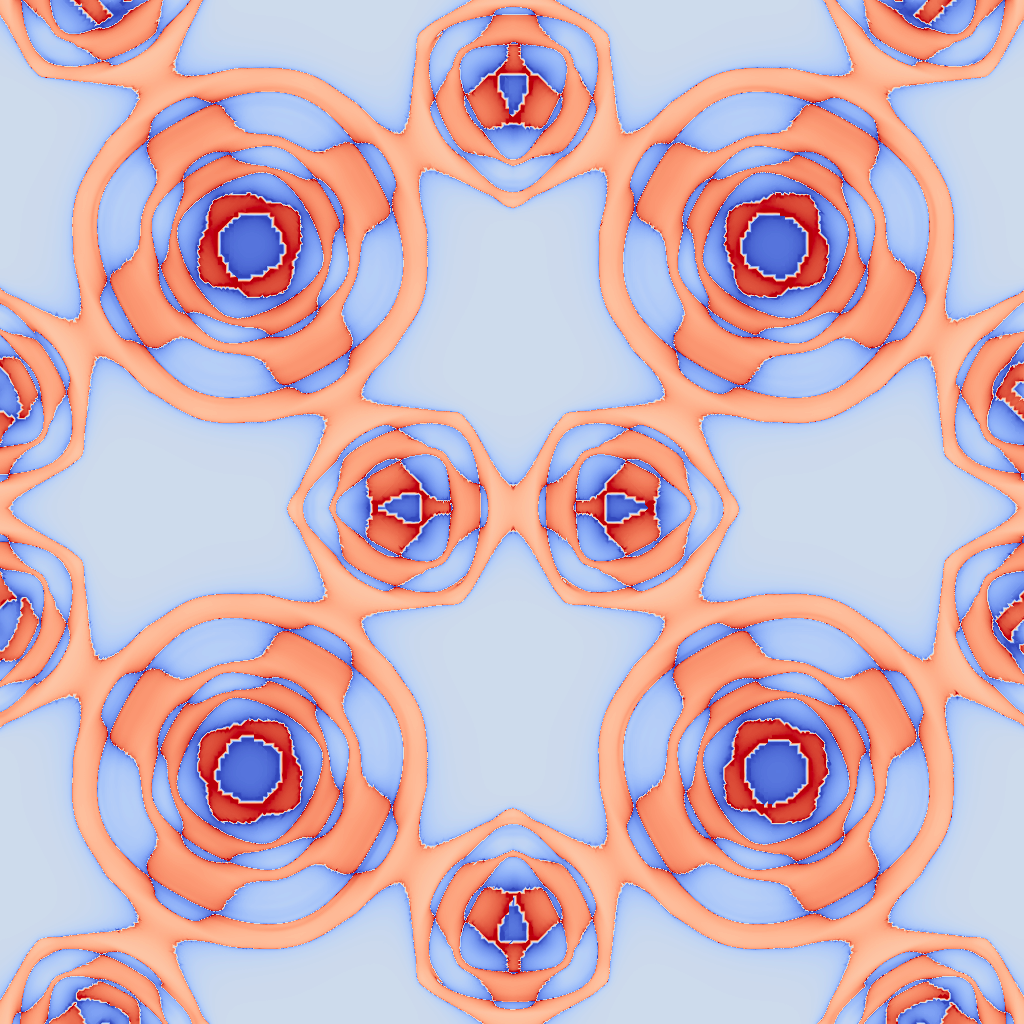}
\includegraphics[width=0.496\linewidth]{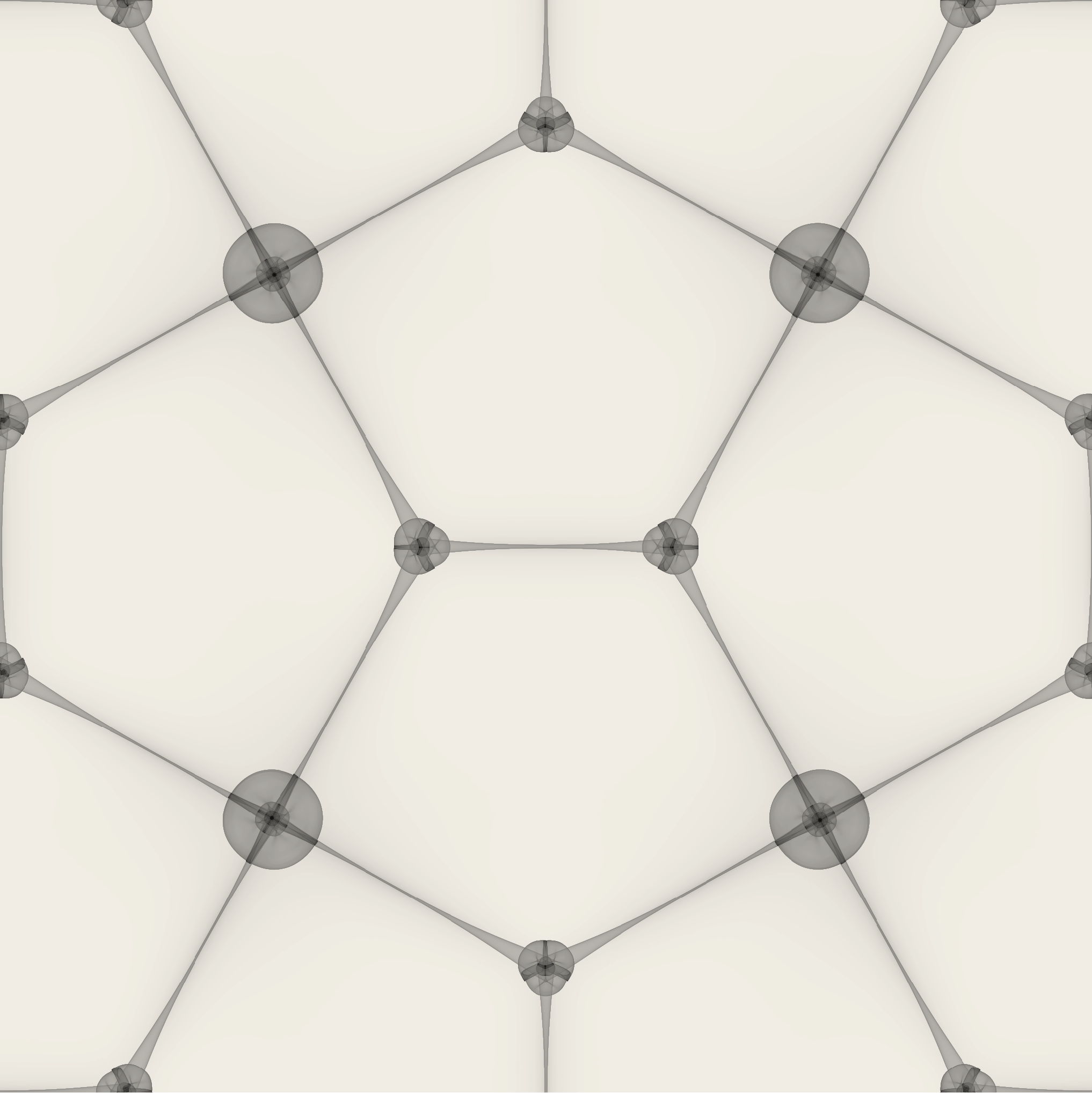}
\caption{A periodic simulated 2D universe that develops from an initial state of circularly symmetric density peaks on vertices of a Cairo pentagonal tiling. {\it Left:} The `crease pattern.' 
Five-sided stars at left only stretch out but do not fold; they become the pentagonal single-layer voids at right. Shading depends on density: dark regions contract during folding; light regions expand. Creases are at sharp light or dark transitions between patches, and delineate regions of positive and negative parity, `face-up' or `face-down' after folding \cite{Neyrinck2012}. {\it Right:} The density field after folding, rendered as though a backlit origami tessellation. The color scale is logarithmic; the cores of nodes are over 1000 times denser than the background.}
\label{fig:cairo_universe}
\end{figure}

How closely does structure formation resemble origami? In a simple, idealized case, rather well. Figure \ref{fig:cairo_universe} shows what a periodic 2D universe would look like if galaxies\footnote{We ignore the important cosmological distinction between a collapsed node of dark matter, called a halo, and a galaxy, which is large enough to host stars.} were arranged on the vertices of an unnaturally regular tiling.  The initial conditions for this simulation were a set of circularly symmetric, smooth density peaks on the vertices of a Cairo pentagonal tiling. We use the ColDICE \cite{SousbieColombi2016} code, a cosmological gravity code which explicitly uses a dark-matter-sheet approach, adaptively refining the resolution of the sheet where necessary.

\begin{figure}
\centering
\includegraphics[width=\linewidth]{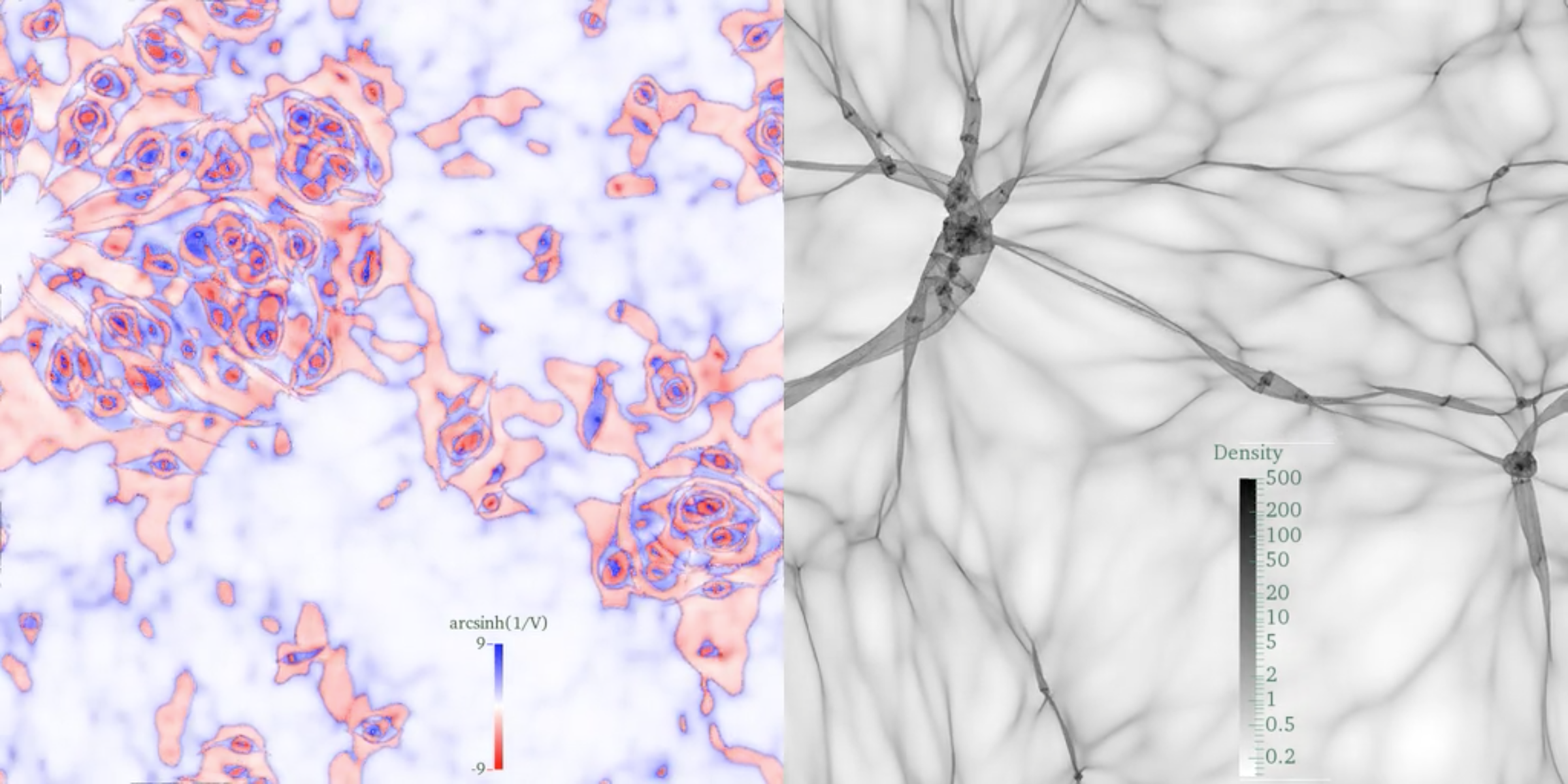}
\caption{A 2D universe in a more realistic, random situation. At left, as in Figure \ref{fig:cairo_universe}, the quantity in the color scale is a nonlinearly-scaled directed density, $\sinh^{-1}(1/V)$, where $V$ is the comoving directed (possibly negative) area of an initial patch, in units of its initial area. Color-scale extremes, \chg{showing crease locations}, have density in units of the mean of $\pm \sinh(9)\sim 400$.}
\label{fig:randomics}
\end{figure}

The outer caustics around each node bear some similarity to an origami tessellation of twist folds. At left, 3-filament nodes are rounded triangles, while 4-filament nodes are rounded squares. Importantly, gravity built these shapes, as it did the filaments around them; in the initial conditions, all nodes had completely circularly symmetric density profiles. There is a fascinating alternating pattern of shapes that comprise the inner structure of each node; the outer crease alternates inward with a stellated version of it. In astronomy, the inner structure of a galaxy is often assumed to be rather independent of its surroundings. But this shows how neighboring structures influence each galaxy; each inner pattern is entirely produced by the external pattern of neighboring nodes around it.

The flattened edges of node outer caustics meet the filaments connecting nodes at 90$^\circ$, as in the Minkowski sum at bottom right of Figure \ref{fig:cosmicduals}. The 90$^\circ$ angle produces a 180$^\circ$ rotation (in 2D, a pure reflection) of the node. In many similar experiments, substantial departures from this 90$^\circ$ angle were very rare. \chg{One obvious way to impart an origami-style `twist' to a node in a cosmological simulation would be to insert a spinning vortex into the initial conditions there. Doing so did cause a few-degree departure from perpendicularity in the simulation, but only with an unphysically strong vortex.}

The crease pattern shown cannot be folded in 3D, without the paper colliding with itself. But it is foldable from a collisionless dark-matter sheet, happy to collide with itself in 3D position space. It can likewise be folded from stretchy paper without collisions in 4D, the position-velocity phase space for a 2D universe.

This tessellation, with possibly appealing regularity, can help to clarify concepts, but the real universe is not so contrived. Figure \ref{fig:randomics} shows a more realistic situation, generated from a random periodic set of initial conditions typical in a 2D slice of the actual universe. Some properties of Figure \ref{fig:cairo_universe} are there, but less clear: alternating patterns of concentric shapes are present inside many nodes, and many filaments form. But the situation is much more irregular and warty. Note, however, that the complicated structure is mostly within the outer caustics, which the adhesion model does not consider; the `cosmic spiderweb' picture is confined to the structure outside these boundaries.

\section{Origami tessellation design}
\label{sec:origamidesign}

\begin{figure}[H]
	\begin{center}
	    \includegraphics[width=0.43\columnwidth]{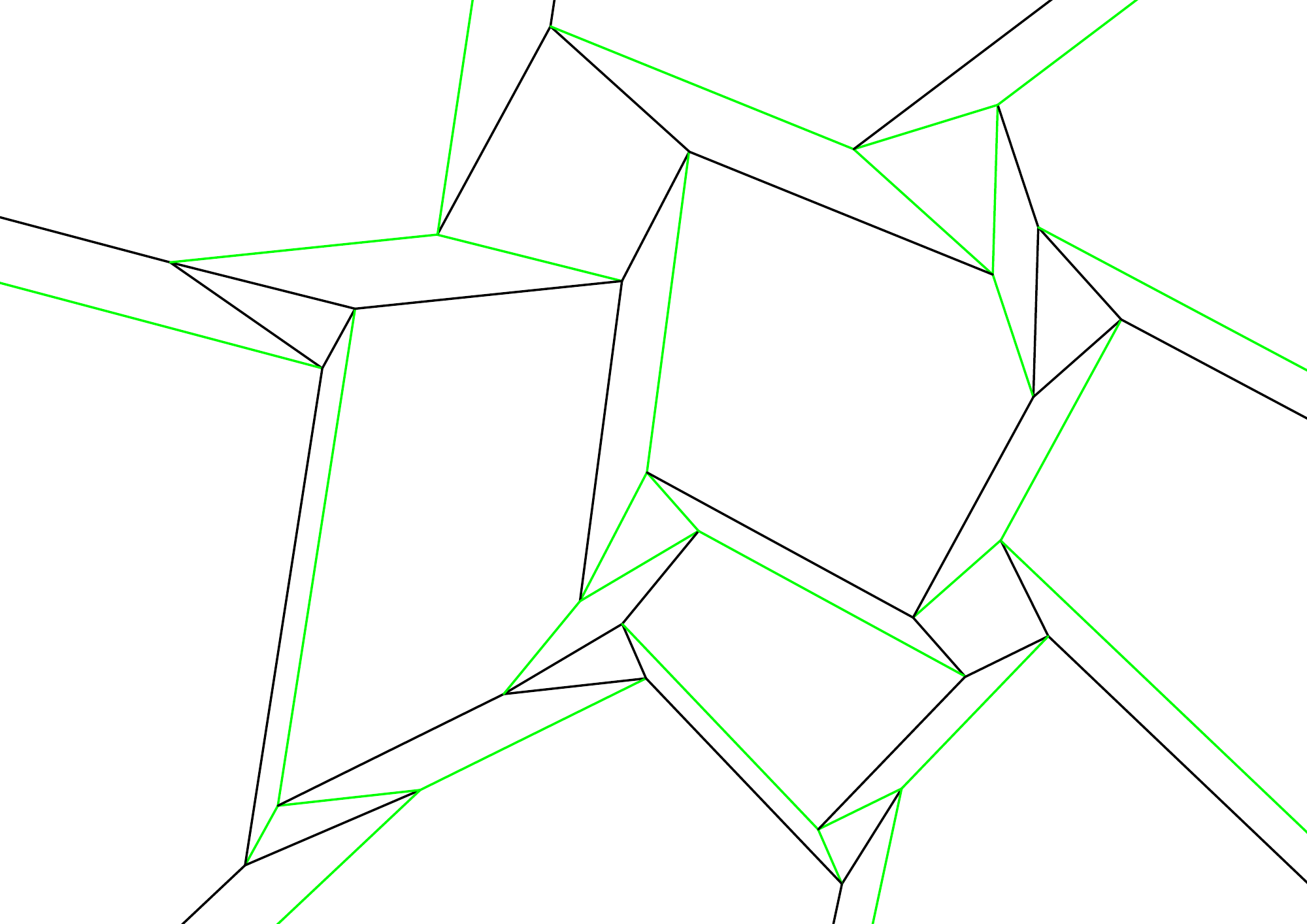}
	    \includegraphics[width=0.56\columnwidth]{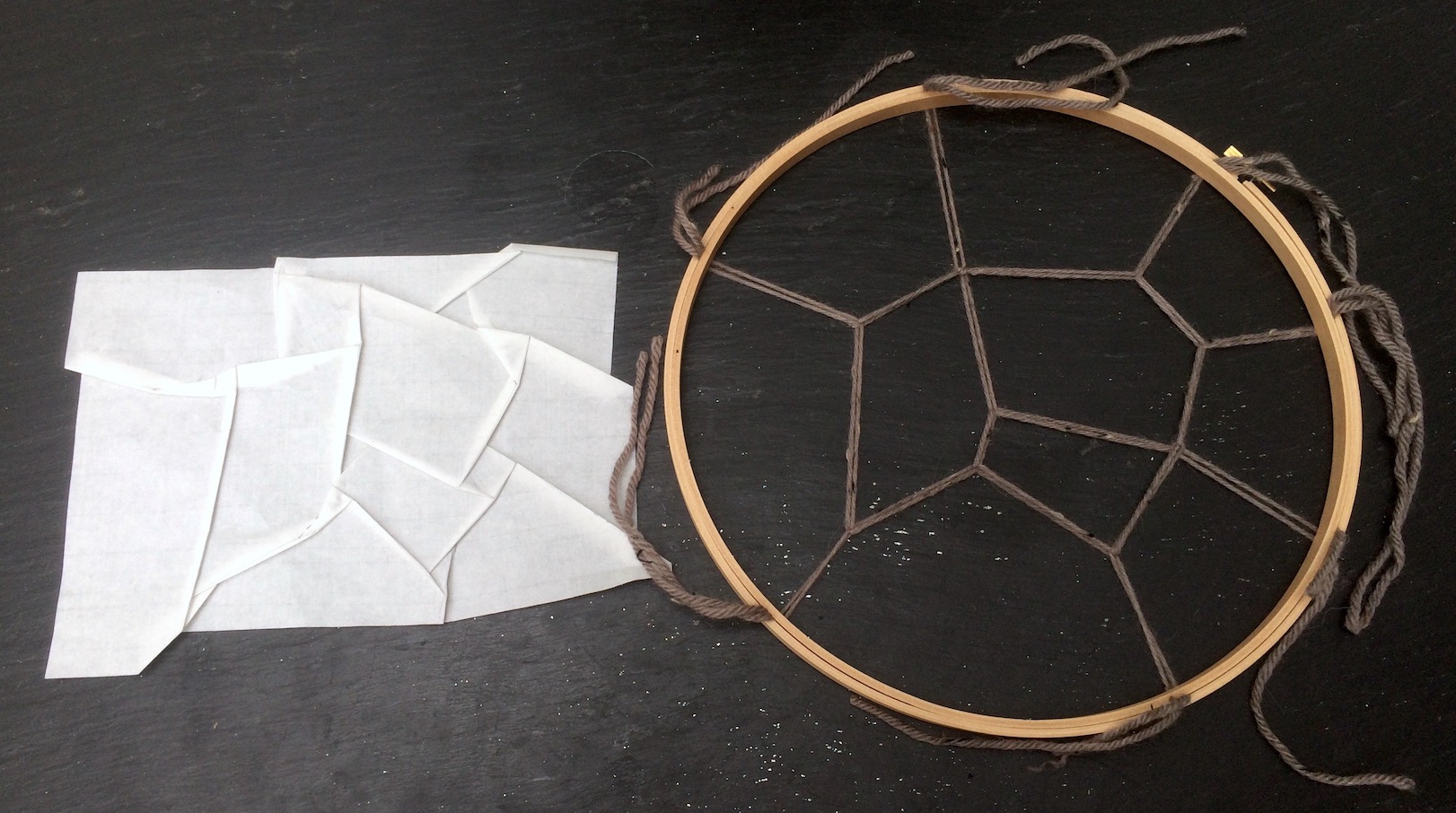}
    	\end{center}
  \caption{Origami and textile approximations of the structure formed by the nearest $\sim$ dozen large galaxies. The Milky Way is nearest the center, Andromeda just under it. {\it Left}: An origami-tessellation crease pattern designed with a Voronoi tessellation. Green and black lines are valley and mountain folds, respectively. {\it Right}: The middle panel folded from paper, alongside a nearly matching spiderweb construction built from yarn and an embroidery hoop. An origami pleat width is proportional to the tension in that strand of the spiderweb.}
  \label{fig:councilofgiants}
\end{figure}

Origami tessellations can be generated from an arbitrary spiderweb \cite{LangBateman2011,Lang2015,lang2017twists}. Robert Lang gives algorithms to generate an origami tessellation from Voronoi and Delaunay tessellations in {\it Tessellatica}\footnote{\hrefurl{http://www.langorigami.com/article/tessellatica}}. These algorithms originated with the `shrink-and-rotate' algorithm for origami tessellation design \cite{Bateman2002}.

Figure \ref{fig:councilofgiants} shows an origami tessellation designed by hand from a Voronoi tessellation in {\it Tessellatica}, and a dreamcatcher-like textile representation of the same geometry. This approximates the structure of the `Council of Giants' \cite{McCall2014}, a term for the nearest dozen or so galaxies of comparable size to the Milky Way. They happen to have a flat arrangement, convenient for 2D representation. Farther out, the arrangement of galaxies becomes fully 3D.

Having developed sectional-Voronoi algorithms in Python for cosmic-web research, it was straightforward to develop `sectional-tess,' a Python package to interactively design sectional-Voronoi tessellations, and therefore spiderwebs and origami tessellations. The package is at \href{https://github.com/neyrinck/sectional-tess}{https://github.com/neyrinck/sectional-tess}, runnable online at \href{https://mybinder.org/v2/gh/neyrinck/sectional-tess/master}{https://mybinder.org/v2/gh/neyrinck/sectional-tess/master}. The package currently lacks important features in e.g.\ {\it Tessellatica}: mountain/valley fold assignment, and checks on foldability. But interactivity, and the flexibility to tune the `power' (additive weight) in each cell, make sectional-tess interesting for origami-tessellation design. In fact, to our knowledge, this flexibility allows a completely arbitrary origami tessellation to be constructed this way, with the important restriction of identical twist angles at every node.
\begin{figure}[H]
  \centering
    	\includegraphics[width=\linewidth]{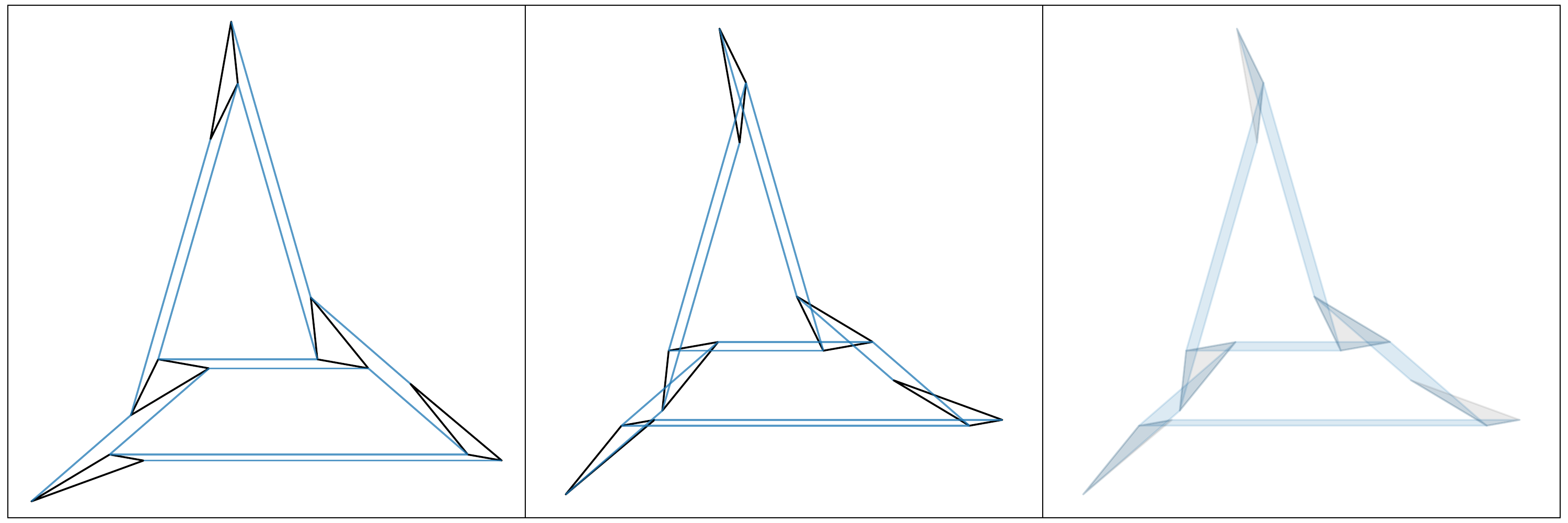}
  \caption{An origami tessellation based on the Delaunay-Voronoi dual similar to Figure \ref{fig:eiffeltower}. {\it Left:} Crease pattern, with no mountain-valley crease assignment. {\it Center:} Folded form, showing only the location of creases. {\it Right:} Folded form, rendered with transparent polygons to indicate layers of paper after folding. For clarity, the largest polygons are left unrendered.}
  \label{fig:origameiffel}
\end{figure}
Figure \ref{fig:origameiffel} shows an origami design based on the schematic Eiffel Tower in Figure \ref{fig:eiffeltower}. The crease pattern is a Minkowski sum, using $\alpha=0.5$, and rotating vectors of the weighted Delaunay tessellation (black, force diagram) by 80$^\circ$ from perpendicularity to the grey, form diagram.

\chg{As far as I am aware, an open question is to characterize the geometry of an arbitrary origami tessellation, with different twist angles at each node. Perhaps there exists an algorithm to go from a spiderweb, plus a twist angle specified at each node, to a unique origami tessellation. A simple generalization of the shrink-rotate algorithm is to assign opposite directions to each adjacent pair of twist folds, keeping the angles constant. In this process, parallelogram pleats in the crease pattern become isosceles trapezoids. This also requires reshaping polygons of the Voronoi/form diagram; alternating sides get shortened or lengthened, as in e.g.\ the `tiled hexagons' design of \cite{Gjerde2008}. This alternation is useful if it is desired that the twisted polygon is on top after folding.}

\chg{Importantly, this alternating-direction crease pattern has a restriction: the spiderweb must be bipartite, i.e. there can be no polygons in the Voronoi/form diagram with an odd number of vertices, since alternating around vertices of each polygon must return to the same parity begun with. It would be fascinating if this restrictions such as these (in 3D, however) applied to some degree in the cosmos, illuminating the way nearby galaxies spin in relation to each other.}

\section{Conclusions}
\label{sec:conclusions}
We discussed shared geometric properties of the cosmic web, spider's webs, and origami tessellations. Other networks visually resemble these; circulatory and road networks, for instance \cite{West2017}. These networks are optimized for efficient transport, which are likely relatable to sectional-Voronoi tessellations, as well, but for totally different reasons. This shared geometry is likely what gives origami tessellations much of their aesthetic appeal. Geometry, the earliest formal mathematical field, continues its usefulness in understanding and appreciating the universe.

\section*{Acknowledgments}
MN thanks Johan Hidding for crucial explanations and illustrative adhesion example code, and Marina Konstantatou and Rien van de Weygaert for contributions to our recent paper.

\bibliographystyle{osmebibstyle}
\small
\bibliography{refs}

\theaffiliations

\end{document}